\documentclass[9pt,conference]{IEEEtran}
\usepackage{dcase2025}
\usepackage{subcaption} 
\usepackage{float}
\usepackage{stfloats}
\usepackage{graphicx}
\usepackage{cuted}
\usepackage{url}
\usepackage{caption} 
\usepackage{afterpage}
\usepackage{booktabs} 
\usepackage{multirow} 
\usepackage{siunitx}  

\usepackage{bm} 

\usepackage{dcase2025,amsmath,graphicx,url,times,booktabs, tabularx}

\title{Importance-Weighted Domain Adaptation for Sound Source Tracking}

\name{Bingxiang Zhong$^{1}$,
      Thomas Dietzen$^{1}$
     \thanks{This project has received funding from KU Leuven internal funds STG/24/013.}}
\address{$^{1}$KU Leuven, Dept. of Electrical Engineering (ESAT-PSI), EAVISE, Sint-Katelijne-Waver, Belgium. \\
 \{bingxiang.zhong, thomas.dietzen\}@kuleuven.be
}

\begin{document}

\maketitle

\begin{abstract}
In recent years, deep learning has significantly advanced sound source localization (SSL). However, training such models requires large labeled datasets, and real recordings are costly to annotate in particular if sources move. While synthetic data using simulated room impulse responses (RIRs) and noise offers a practical alternative, models trained on synthetic data suffer from domain shift in real environments. Unsupervised domain adaptation (UDA) can address this by aligning synthetic and real domains without relying on labels from the latter. The few existing UDA approaches however focus on static SSL and do not account for the problem of sound source tracking (SST), which presents two specific domain adaptation challenges. First, variable-length input sequences create mismatches in feature dimensionality across domains. Second, the angular coverages of the synthetic and the real data may not be well aligned either due to partial domain overlap or due to batch size constraints, which we refer to as directional diversity mismatch. To address these, we propose a novel UDA approach tailored for SST based on two key features. We employ the final hidden state of a recurrent neural network as a fixed-dimensional feature representation to handle variable-length sequences. Further, we use importance-weighted adversarial training to tackle directional diversity mismatch by prioritizing synthetic samples similar to the real domain. Experimental results demonstrate that our approach successfully adapts synthetic-trained models to real environments, improving SST performance.

\end{abstract}


\section{Introduction}
\label{sec:intro}
Sound source localization (SSL), which refers to estimating the direction of arrivals (DoAs) of sound sources, is fundamental to applications requiring directional information, including speech enhancement \cite{lee2016dnn}, human-robot interaction \cite{argentieri2015survey}, and automotive applications \cite{schulz2021hearing}. Traditional approaches exploit explicit models based on time difference of arrivals (TDoAs), such as steered response power with phase transform (SRP-PHAT) \cite{srp-phat}, multiple signal classification (MUSIC) \cite{music}, and estimation of signal parameters via rotational invariance techniques (ESPRIT) \cite{roy2002esprit}. However, these methods suffer significant performance degradation in challenging acoustic conditions with low signal-to-noise ratios (SNR) and high reverberation.

Deep learning has recently demonstrated substantial improvements \cite{cross3D, yang2022srp-dnn, neual-srp, grumiaux2022survey} over traditional SSL methods, but their success critically depends on large-scale labeled datasets. Labeling data is time-consuming and costly, especially for moving sources, leading to a scarcity of real-world labeled data. As an alternative, room impulse response (RIR) simulation tools \cite{rir-generator, gpurir} are used to generate extensive synthetic training data with diverse SNR and reverberation conditions, while small portions of real-world data serve for evaluation. However, models trained on synthetic data typically suffer performance drops due to simplified simulation models that do not fully capture the complexity of room acoustics and microphone characteristics, creating distribution mismatches between synthetic and real data, a phenomenon known as domain shift \cite{domain-shift}.

Unsupervised domain adaptation (UDA) attempts to mitigate the effect of domain shift by aligning features across source (synthetic) and target (real-world) domains without requiring target labels. While extensively investigated in computer vision and natural language processing \cite{domain-app-review}, few studies have addressed this problem in SSL. Takeda et al. \cite{takeda2017unsupervised} proposed entropy minimization to regularize DoA classification probabilities, and later \cite{takeda2018unsupervised} introduced an eliminative posterior probability constraint that forces the probability of less likely candidates to become zero to eliminate incoherent errors. However, both approaches are limited to discrete classification formulations. Domain adversarial training \cite{DANN} for SSL has shown mixed results. While \cite{he2019adaptation} reported minimal improvements, \cite{le2021Data-efficient} achieved significant gains using ensemble domain adversarial networks with multiple domain discriminators at different network layers. Notably, none of these approaches address domain adaptation specifically for sound source tracking (SST).

SST presents unique UDA challenges beyond static localization. First, tracking requires processing variable-length temporal sequences, while many existing domain adaptation methods \cite{DANN, coral, domain-app-review} are designed for fixed feature dimensionality and cannot directly handle temporal variability. Second, directional diversity mismatch between source and target data creates feature alignment instability in domain adaptation through two effects. At the domain level, angular coverages may differ significantly. While source domains typically contain full directional diversity, target domains may cover only limited angular ranges depending on recording settings, creating partial domain overlap \cite{zhang2018importance}. At the batch level, SST systems must handle long-duration audio sequences (typically  $>10$ seconds \cite{cross3D, yang2022srp-dnn}), necessitating small batch sizes due to memory constraints. Consequently, directional diversity within batches may differ significantly between source and target data, creating inconsistent feature distributions that hinder effective domain alignment. These challenges render standard adversarial training approaches \cite{le2021Data-efficient, he2019adaptation} ineffective, as they fundamentally assume complete label space alignment both across domains and within batches \cite{zhang2018importance}.

To address these challenges, we propose a novel UDA approach tailored for SST with two key contributions. First, adopting the SST architecture from \cite{yang2022srp-dnn, realmandataset}, we employ the final hidden state of a recurrent neural network (RNN) as a fixed-dimensional feature representation within the domain discriminator to handle variable-length sequences. Second, we leverage importance-weighted adversarial training \cite{zhang2018importance} to tackle directional diversity mismatch by prioritizing synthetic samples similar to the real domain. Experimental results demonstrate that our approach successfully adapts synthetic-trained models to real environments, improving SST performance. The code is available at \cite{code}.

In section 2, we review a commonly used SST model adopted for domain adaptation. In section 3, we  describe the proposed domain adaptation approach. Section 4 presents our experimental evaluation, and section 5 concludes the paper.

\begin{figure}
    \centering
    \begin{subfigure}{0.32\textwidth}
        \centering
        \includegraphics[width=1\linewidth]{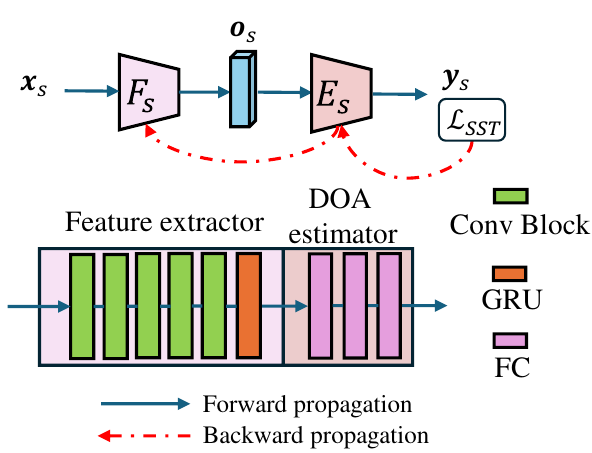}
        \caption{}
        \label{fig:sub1}
    \end{subfigure}
    \hspace{0.2\textwidth}
    \begin{subfigure}{0.35\textwidth}
        \centering
        \includegraphics[width=1\linewidth]{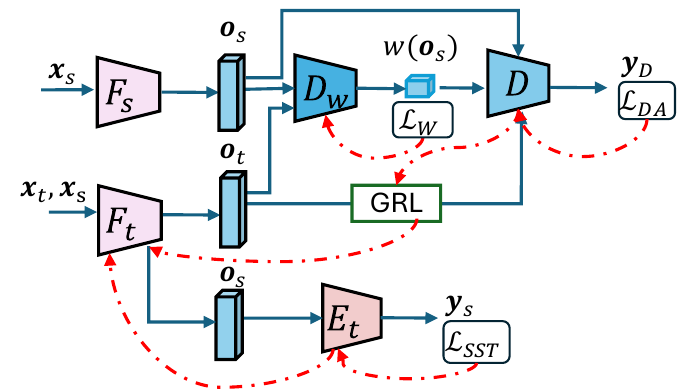}
        \caption{}
        \label{fig:sub2}
    \end{subfigure}
    \caption{The overview of the proposed method. (a) The CRNN model will first be trained on the source domain. (b) The pretrained model is adapted to the target domain using importance-weighted adversarial training. }
    \label{fig:proposed method}
\end{figure}
\vspace{-5pt}

\section{Sound Source Tracking}
We adopt the convolutional recurrent neural network (CRNN) architecture used in \cite{yang2022srp-dnn, realmandataset} to perform single-source SST. The model is first trained on source domain data, considering tracking of sound sources in the azimuth direction spanning the range $[0, \pi)$. The input to the SST system concatenates the real and imaginary components of the short-time Fourier transform (STFT) of multi-channel acoustic signals, denoted as $\mathbf{x}_s$ of size $2M \times K \times L$, where $M$ is the number of microphone channels, $K$ is the number of frequency bins, and $L$ is the number of time frames. Rather than directly regressing angular directions, we employ likelihood encoding as in \cite{he2018deep, realmandataset, he2019adaptation}, where the network output represents the probability of source existence across $J = 180$ candidate directions spanning the range $[0, \pi)$. The model is trained to regress a sequence of likelihood values for ground truth DoA values in the azimuth direction, denoted as $\mathbf{y}_s$ with dimensions $L \times J$. Given ground truth DoA values $\phi^\prime(l)$ at each voice-active time frame $l$,  and 180 candidate directions $\{\phi_i\}_{i=1}^{180}$, the desired output values are encoded via Gaussian functions 
\begin{equation}
y_s(\phi_i, l) = \exp\Big\{-\frac{(\phi_i- \phi^\prime(l))^2}{2\sigma^2}\Big\},
\end{equation}
where $\sigma$ defines the beam width, set to 16$^\circ$ as in \cite{realmandataset}. For non-voice-active frames, we set $y_s(\phi_i, l) = 0$. Voice-active frames are identified using the WebRTC voice activity detector (VAD) \cite{pywebrtcvad}.

The network architecture of the adopted SST model \cite{yang2022srp-dnn, realmandataset} is shown in \cref{fig:sub1}. To facilitate the following explanation of domain adaptation, we split the model into two parts: a feature extractor $F_s$ and a DoA estimator $E_s$, both specifically for the source domain. The feature extractor comprises five convolutional blocks followed by a gated recurrent unit (GRU). Each convolutional block contains two consecutive convolutional layers with 3×3 kernel size and 64 output channels, batch normalization, ReLU activation, and max pooling. Max pooling operations have kernel sizes of [4, 2, 2, 2, 2] along the frequency axis and [1, 1, 1, 1, 5] along the time axis. The resulting convolutional features are processed by a unidirectional GRU with 256 hidden units. The output of the GRU, denoted by $\mathbf{o}_s = F_s(\mathbf{x}_s)$, serves as the input to the DoA estimator $E_s$ and also functions as a domain feature in the proposed domain adaptation approach. The DoA estimator consists of three fully connected (FC) layers with dimensions 512, 256, and 180, utilizing Tanh, ReLU, and sigmoid activations, respectively.

The model is trained to minimize the mean squared error (MSE) between the network predictions and the ground truth DoA likelihood representations. Given a batch of $N$ input data samples $\mathbf{X}_s = \{\mathbf{x}_s^{(n)}\}$ with corresponding labels $\mathbf{Y}_s = \{\mathbf{y}_s^{(n)}\}$, $ n =1,...,N$, the MSE loss for SST is computed as
\begin{equation}
\mathcal{L}_{SST}(F_s, E_s) = \frac{1}{N} \sum_{n=1}^N || E_s(F_s(\mathbf{x}_s^{(n)})) - \mathbf{y}_s^{(n)} ||_2^2,
\end{equation}
where both $F_s$ and $E_s$ are updated during backward propagation. During inference, the estimated DoA is obtained by identifying the peak in the estimated encoded likelihood.

\section{Importance-weighted domain adaptation for sound source tracking}
\label{sec: method}
Let $\mathcal{D}_s = \{\mathbf{X}_s, \mathbf{Y}_s\}$ represent the labeled source domain and $\mathcal{D}_t = \{\mathbf{X}_t\}$ represent the unlabeled target domain, where data samples are drawn from distributions $p_s$ and $p_t$ respectively. When deploying the model trained on the source domain to the target domain, the domain shift of the distribution mismatch between source and target domain data, i.e., $p_s\neq p_t$, leads to performance degradation on the target domain. 

While domain adaptation approaches exist to address such distribution mismatch, SST presents two particular challenges: variable length features and diectional diversity mismatch between $\mathcal{D}_s$ and $\mathcal{D}_t$. To address these issues, we propose to use the final hidden state of the RNN to handle variable length sequences and adopt importance-weighted domain adversarial training \cite{zhang2018importance} originally introduced for image classification and employ it for SST. The resulting architecture is shown in \cref{fig:sub2}.


The adaptation process begins by initializing the target domain feature extractor $F_t$ and DoA estimator $E_t$ with weights from the pretrained model. During domain adaptation, $F_s$ remains frozen to maintain consistent source domain representations, while $F_t$ adapts to extract domain-invariant features from both the source and target domain data. The approach incorporates two binary domain discriminators, $D_w$ and $D$, which share an identical architecture. Source samples are labeled as 1 and target samples as 0. $D_w$ facilitates the importance weighting mechanism, while $D$ performs domain adversarial training. In the following, we describe the domain discriminator architecture, importance-weighted adversarial training, and importance weighting mechanism.



\subsection{Domain discriminator architecture}

To handle the variable-length input sequences, we design our discriminator architecture to effectively aggregate long and variable-length temporal dynamics. Unlike previous works \cite{he2019adaptation, le2021Data-efficient} that operate on fixed-dimensional features and can therefore use convolutional layers or simple FC layers for domain discrimination, our approach must address the challenge of variable-length temporal features. Therefore, we design our discriminator with a GRU and employ its final hidden state to encode the complete temporal dynamics into a fixed-dimensional representation independent of input length.

Both discriminators first process features $\mathbf{o}_s$ and $\mathbf{o}_t$ from the feature extractors through a 256-unit GRU layer. The final hidden state of this GRU then passes through two FC layers: a 128-unit layer with ReLU activation followed by a single-output layer with sigmoid activation for domain classification. 
\subsection{Importance-weighted adversarial training}
The importance-weighted adversarial training \cite{zhang2018importance} follows the same training approach as standard adversarial training \cite{DANN}. The core principle is to align feature representations across domains through a weighted adversarial game between a feature extractor and a domain discriminator. In our implementation, the features $\mathbf{o}_s$ and $\mathbf{o}_t$ extracted from source and target domain data serve as input to the domain discriminator $D$. The loss function is given by \cite{zhang2018importance},
\begin{equation}
\begin{split}
\mathcal{L}_{DA}(D, F_t) &= \frac{1}{N} \sum_{n=1}^N  [ w(\mathbf{o}^{(n)}_s)\log D(\mathbf{o}^{(n)}_s) \\ &
 +  \log(1 - D(\mathbf{o}^{(n)}_t))], \\
\end{split}
\label{eq:minimax}
\end{equation}
where $\mathbf{o}^{(n)}_s = F_s(\mathbf{x}^{(n)}_s)$ and $\mathbf{o}^{(n)}_t = F_t(\mathbf{x}^{(n)}_t)$. For standard adversarial training \cite{DANN}, $w(\mathbf{o}^{(n)}_s) = 1$. The objective is to solve a minimax game, $\min_{F_t} \max_{D} \mathcal{L}_{DA}(D, F_t)$. The discriminator $D$ maximizes the objective by correctly classifying source domain features and target domain features. Simultaneously, the target feature extractor $F_t$ minimizes the objective by generating features that fool the discriminator. Through this adversarial process, the feature extractor learns domain-invariant representations that enable effective knowledge transfer from the labeled source domain to the unlabeled target domain. A gradient reversal layer (GRL) enables simultaneous training of the minmax game by reversing gradients for the feature extractor. 
\subsection{Importance weighting mechanism}
To obtain the importance weights $w$, a second domain discriminator $D_w$ is employed. The goal is to assign larger weights to source data samples that share similar feature characteristics with the target data, while assigning smaller weights to those with less overlap. This discriminator is first trained to distinguish between source and target domains using the binary cross entropy loss,
\begin{equation}
\mathcal{L}_{W} (D_w) = -\frac{1}{N} \sum_{n=1}^N [\log D_w(\mathbf{o}^{(n)}_s) 
 + \log(1 - D_w(\mathbf{o}^{(n)}_t))].
\end{equation}
Since the final activation function in $D_w$ is the logistic sigmoid function, the output $D_w(\mathbf{o}_s)$ represents the probability that a source sample belongs to the source domain. When $D_w(\mathbf{o}_s) \approx 1$, source samples are highly distinguishable from target samples and should receive lower weights in the adversarial training. Conversely, when $D_w(\mathbf{o}_s) \approx 0$, the source samples are similar to target samples and should receive higher weights. The importance weight is therefore inversely related to $D_w(\mathbf{o}_s)$,
\begin{equation}
w(\mathbf{o}_s) = 1-D_w(\mathbf{o}_s).
\end{equation}
These weights are normalized by the mean weight value per batch \cite{zhang2018importance}. Note that the domain discriminator $D_w$ is used solely to calculate the importance weights and does not participate in the adversarial training; therefore, gradients from $D_w$ do not back propagate to update the feature extractor $F_t$. The obtained weights are then incorporated into the loss function in \cref{eq:minimax} to prioritize source samples that are most similar to the target domain during adversarial training.


\subsection{Overall training objective}
To preserve source domain performance while adapting to the target domain, we minimize the combined loss function,
\begin{equation}
\begin{split}
    \mathcal{L} (D, D_w, F_t, E_t) &= \mathcal{L}_{SST}(F_t, E_t)  \\
   & +\lambda \mathcal{L}_{DA}(D, F_t) +\mathcal{L}_{W}(D_w).\\
\end{split}
\end{equation}
Here, $\mathcal{L}_{SST}$ is computed exclusively using source domain data to retain performance on the original task.
The trade-off parameter $\lambda$ controls the influence of the domain adversarial loss. When $\lambda$ is too small, domain alignment is weak, and the model behaves similarly to source-only training. When $\lambda$ is too large, the adversarial objective dominates, encouraging excessive domain invariance and causing the learned features to lose essential task-relevant characteristics.

\section{Experiments}
\label{sec: exp}

\subsection{Experimental setups}
\subsubsection{Datasets}
\textbf{Target Domain (Real-world Data)}: we use the RealMAN dataset \cite{realmandataset}, which contains 35.4 hours of moving speaker recordings across 32 different acoustic scenes with background noise, sampled at 48 kHz. The dataset employs a 32-channel microphone array, from which we extract a 9-channel subarray consisting of an 8-channel uniform circular array (3 cm radius) plus a center microphone.
In the experiments, we consider single-source moving scenarios in three distinct acoustic scenes spanning indoor to outdoor environments: OfficeRoom3, OfficeLobby, and BasketballCourt1. These scenes provide training, validation, and test data with durations of (81, 7, 8), (51, 28, 8), and (40, 8, 42) minutes, respectively. We select these three scenes because they provide complete training, validation, and testing splits along with scene-specific background noise recordings. During adversarial training, the training data is augmented with background noise at SNR levels ranging from -10 dB to 15 dB, matching the SNR range found in the validation and test data \cite{realmandataset}.

 \textbf{Source Domain (Synthetic Data)}: we generate synthetic data using the same 9-channel microphone configuration as the target domain to ensure acoustic consistency. Following established protocols \cite{cross3D}, we simulate rooms with dimensions randomly selected from 3×3×2.5 m to 10×8×6 m and reverberation times from 0.2 s to 1.0 s. RIRs are generated using gpuRIR \cite{gpurir}.
Speech signals are randomly selected from the LibriSpeech corpus \cite{librispeech} and convolved with the generated RIRs, then corrupted with diffused noise at SNRs ranging from $-10$ dB to 15 dB. The diffused noise is generated using the toolbox \cite{diffuisenoisegenerating} applied to the NOISEX-92 corpus \cite{noisex-92}. Speech segments vary from 2 s to 10 s in duration. To enhance training diversity, each sample is generated on-the-fly with randomized acoustic conditions including source trajectories, microphone positions, source signals, noise characteristics, reverberation times, and SNRs.

Both target and source domain data is downsampled to 16 kHz for computational efficiency. STFT analysis uses a 32 ms window with 20 ms frame shift to extract time-frequency representations.

\subsubsection{Evaluation metrics}

Performance is evaluated during voice-active segments using mean absolute error (MAE) of azimuth estimates for across all time frames, and localization accuracy (Acc), defined as the percentage of frames with MAE below 5$^\circ$.

\subsubsection{Training configuration}

The training process comprises two phases. In the first phase, the model is trained on the source domain for 150 epochs with a batch size of 16, using the Adam optimizer \cite{adam} and a learning rate of 0.0001. The model checkpoint is selected based on its MAE performance on simulated testing data.

Subsequently, the model undergoes domain adaptation for an additional 60 epochs with the same learning rate, and early stopping. To mitigate training instability in early stages, we employ a gradual schedule for the adversarial trade-off parameter following \cite{DANN},
\begin{equation}
    \lambda = \frac{2u}{1+\exp(- p)} - u,
    \label{upper bound}
\end{equation}
where $p$ gradually increases over training steps, allowing $\lambda$ to smoothly approach its upper bound $u$. The model checkpoint selection is also based on the model's MAE performance on a small validation set. Although this may appear to conflict with the principles of unsupervised learning, validation-based model selection is widely accepted in domain adaptation for computer vision tasks \cite{DANN, coral, zhang2018importance}. While alternative strategies for checkpoint selection without validation sets exist \cite{evalDAwithout}, they are typically tailored to classification problems. Moreover, collecting and annotating a small validation set is far more feasible than creating large annotated training datasets.

\subsubsection{Baselines}
We denote the proposed method as SST-IWDA. To evaluate its effectiveness, we compare it with two baseline models. The first baseline, SST-SO, is a source-only model trained exclusively on source domain data. For a fair comparison, the pretrained model is further trained for 60 additional epochs, with model checkpoint selection based on validation performance on the target domain data for each acoustic scene. The second baseline, SST-DA, implements standard adversarial domain adaptation \cite{DANN} without the importance weighting mechanism.

\subsection{Results and discussion}
\begin{figure*}[t]
    \centering
    \subfloat[OfficeRoom3\label{fig:baseline}]{%
        \includegraphics[height=0.2\textwidth, width=0.25\textwidth]{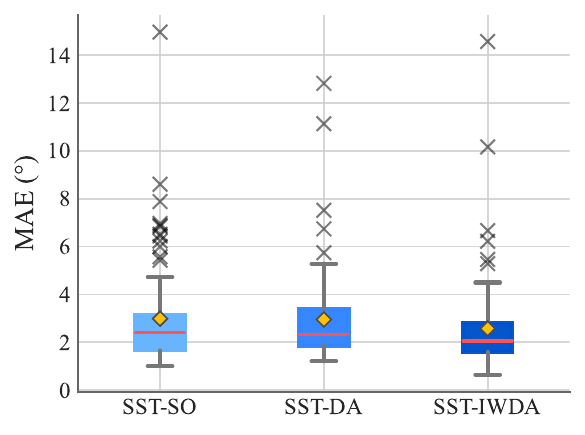}
    }\hspace{0.05\textwidth}
    \subfloat[OfficeLobby\label{fig:dann}]{%
        \includegraphics[height=0.2\textwidth, width=0.25\textwidth]{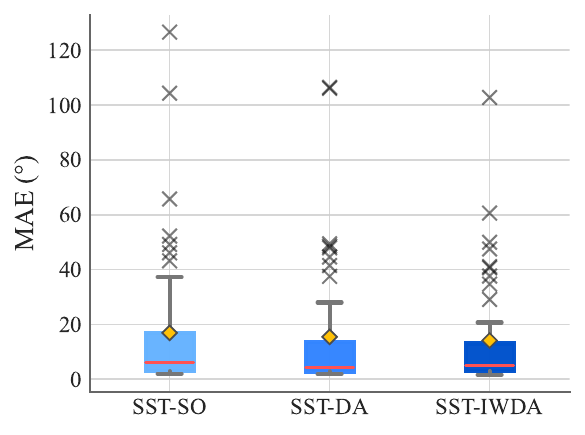}
    }\hspace{0.05\textwidth}
    \subfloat[BasketballCourt1\label{fig:proposed}]{%
        \includegraphics[height=0.2\textwidth, width=0.25\textwidth]{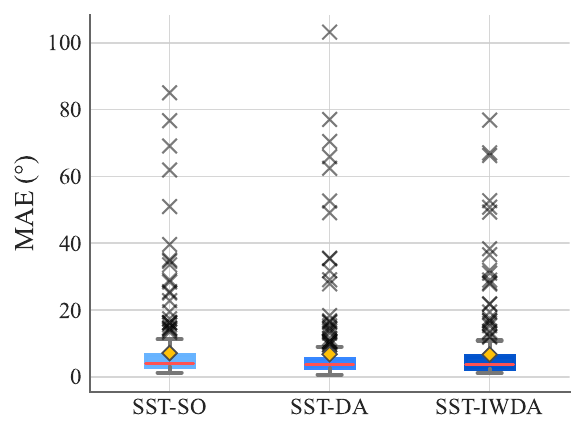}
    }

    \caption{Boxplots comparing the MAE distributions of different methods across various acoustic scenes. The boxes span the interquartile range (IQR, 25th to 75th percentiles), with the red line marking the median. Whiskers extend to the minimum and maximum values within 1.5×IQR from the quartiles. Diamond markers indicate mean MAE values, while crosses denote outliers beyond the whiskers.}
    \label{fig:boxplot}
\end{figure*}

In the first experiment, we adapt the pretrained model using target data from various acoustic environments. The results using $u=0.001$, presented in \cref{tab:results}, show that SST-DA yields modest performance gains over the source-only baseline SST-SO. However, the proposed approach SST-IWDA consistently achieves the lowest MAE and highest accuracy among all methods. For example, in OfficeRoom3, SST-IWDA reduces the MAE from 3.03$^\circ$ to 2.57$^\circ$ and boosts accuracy from 80\% to 89\%. Similarly, in BasketballCourt1, we observe a reduction in MAE from 6.26$^\circ$ to 5.62$^\circ$, with accuracy improving from 68\% to 76\%.


While SST-SO performs reasonably well in OfficeRoom3 and BasketballCourt1, both characterized by relatively low MAEs, its performance deteriorates substantially in OfficeLobby, where the estimation error reaches 17.12$^\circ$. This degradation is likely due to a pronounced mismatch in spatial correlation coefficients between the real-world office lobby noise and the diffuse noise assumption employed in our simulated training data \cite{realmandataset}. Despite this domain gap, SST-IWDA improves the MAE by nearly 2$^\circ$ (from 17.12$^\circ$ to 15.39$^\circ$) and increases accuracy by 5 percentage points.
\begin{table}
\centering
\caption{Performance comparison across different acoustic scenes (lower MAE and higher Acc indicate better performance).}
\label{tab:results}
\scalebox{0.85}{ 
\begin{tabular}{@{}l 
                S[table-format=1.2,detect-weight] 
                S[table-format=2.0,detect-weight]
                S[table-format=1.1,detect-weight] 
                S[table-format=2.0,detect-weight]
                S[table-format=1.2,detect-weight] 
                S[table-format=2.0,detect-weight]@{}}
\toprule
\multicolumn{1}{c}{} &
\multicolumn{2}{c}{\textbf{OfficeRoom3}} & 
\multicolumn{2}{c}{\textbf{OfficeLobby}} & 
\multicolumn{2}{c}{\textbf{BasketballCourt1}} \\
\cmidrule(lr){2-3} \cmidrule(lr){4-5} \cmidrule(l){6-7}
\cmidrule(lr){2-3} \cmidrule(lr){4-5} \cmidrule(l){6-7}
\multicolumn{1}{@{}l}{\textbf{Method}} &
\multicolumn{1}{c}{MAE (\si{\degree})} & \multicolumn{1}{c}{Acc (\%)} & 
\multicolumn{1}{c}{MAE (\si{\degree})} & \multicolumn{1}{c}{Acc (\%)} & 
\multicolumn{1}{c}{MAE (\si{\degree})} & \multicolumn{1}{c}{Acc (\%)} \\
\midrule
SST-SO & 3.03 & 80 & 17.12 & 51 & 6.26 & 68 \\
SST-DA & 2.92 & 83 & 16.19 & \textbf{56} & 5.93 & 71 \\
SST-IWDA & \textbf{2.57} & \textbf{89} &\textbf{15.39} & \textbf{56} & \textbf{5.62} & \textbf{76} \\
\bottomrule
\end{tabular}}
\end{table}

The boxplot analysis in \cref{fig:boxplot} reveals additional insights into the performance of the proposed approach. Domain adaptation can reduce both the number of outliers and their magnitude, which correlates with the observed increase in accuracy. SST-IWDA consistently reduces the interquartile range (IQR) and whisker lengths, as well as the median values. In contrast, SST-DA exhibits inconsistent behavior: it sometimes increases the quartile values and whisker lengths (as observed in OfficeRoom3), while in other cases it reduces the IQR effectively but introduces outliers with larger magnitudes (as seen in BasketballCourt1).

\begin{figure}
    \centering
    \subfloat[MAE]{\includegraphics[width=0.48\linewidth]{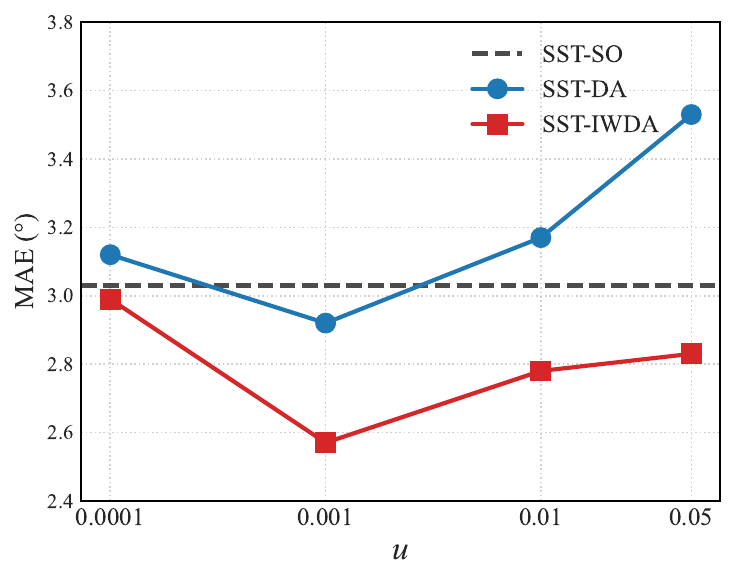}\label{fig:mae}}
    \hfill
    \subfloat[Acc]{\includegraphics[width=0.48\linewidth]{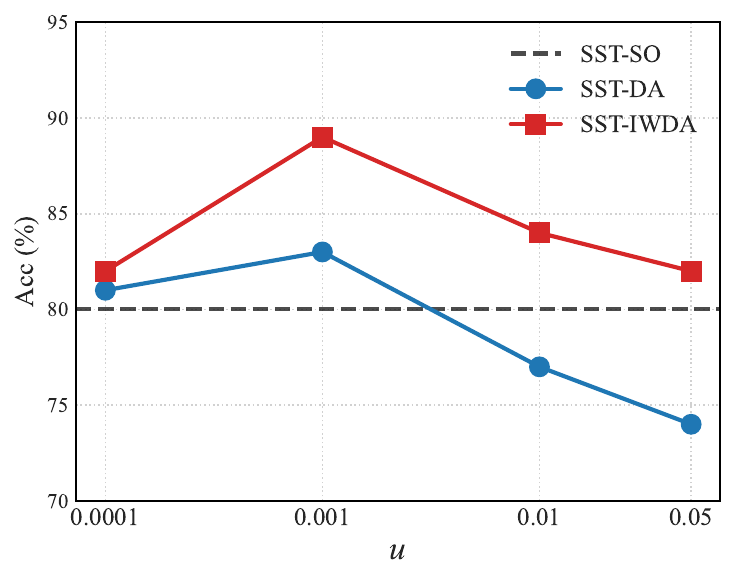}\label{fig:acc}}
    \caption{Domain adaptation performance across different values of parameter $u$ for the OfficeRoom3 acoustic scene. }
    \label{fig:upper bound}
\end{figure}

In the second experiment, we evaluate domain adaptation performance in OfficeRoom3 under varying values of the upper bound parameter $u$ from \cref{upper bound}, which controls the strength of adversarial training. We test $u$ of values 0.0001, 0.001, 0.01, 0.05 for both SST-DA and SST-IWDA.
As shown in \cref{fig:upper bound}, small $u$ values yield performance similar to SST-SO. Increasing $u$ initially improves results, but excessive values degrade both MAE and accuracy. SST-IWDA consistently outperforms SST-SO across all tested values, whereas SST-DA rapidly surpasses SST-SO’s MAE (3.03$^\circ$) and drops below its Acc (80\%) as $u$ rises. These findings confirm the robustness of the proposed importance-weighted adversarial training in balancing adaptation and task-specific learning.

\section{Conclusion}
This paper proposes a novel domain adaptation approach for SST with two key contributions. First, we utilize GRU's final hidden state to encode variable-length sequences into fixed-dimensional representations within the domain discriminator. Second, we use importance-weighted adversarial training to mitigate directional diversity mismatch between synthetic and real data distributions. Experimental results demonstrate that the proposed method outperforms baseline approaches and exhibits enhanced robustness across different scenes.

\clearpage
\bibliographystyle{IEEEtran}
\bibliography{refs}







\end{document}